# Electron Transport and Thermoelectric Performance of Defected Monolayer MoS$_2$

Munish Sharma[1, a)], Ashok Kumar[2] and P.K.Ahluwalia[1]

[1)] *Department of Physics, Himachal Pradesh University, Shimla 171005, India*

[2)] *Center for Physical Sciences, School of Basic and Applied Sciences, Central University of Punjab, Bathinda,India, 151001*

(March 08, 2017)

[a)]Corresponding Authors

Email: munishsharmahpu@live.com

# Abstract

Electronic and thermoelectric properties of a two-dimensional $MoS_2$ monolayer containing atomic defects are investigated using density functional theory. All the atomic defects have been found to exhibit endothermic nature. Electronic structure of $MoS_2$ shows tuneability of band gap with the atomic defects. The $MoS_2$ vacancy in pristine monolayer makes it magnetic and narrow band gap semiconductor. The spin-polarized character of the monolayer with defects is clearly captured by the tunneling current calculated in the STM-like setup. A relatively low thermal conductivity has been observed in monolayers with defects as compared to pristine form resulting in enhanced room temperature figure of merit as high as 6.24 and 1.30 respectively. We believe that our results open up a new window for the use of monolayer $MoS_2$ in electronic devices, thermal management and thermoelectric devices.

## 1. Introduction

The discovery of graphene [1] opened a flood gate to the world of two dimensional materials. The award of noble prize to A. K. Geim and K. Novoselov in 2010 clearly acknowledge the novelty of graphene as a material of immense possibilities [2-5]. However, absence of inherent band gap in graphene hindered its direct applications. The other two dimensional (2D) alternatives to graphene such as $MoS_2$ and other few-layer transition metal dichalcogenides (TMDs) are offering opportunities for energy conversion [6], sensing [7-10], super lubricity in nano-machines [11], photoelectronic application [12] and electronic device applications [13-15].

Monolayer $MoS_2$ could be synthesized by mechanical exfoliation technique [16], liquid exfoliation [17] and chemical synthesis techniques such as chemical vapor deposition (CVD) [18,19], physical vapor deposition (PVD) [20] etc. Note that a defect-free structure is not a practical reality as atomic defects have been observed in experiments [21]. Recent investigations suggest that the carriers mobility decreases in samples prepared by chemical methods as compared to samples prepared by mechanical exfoliation technique [18,22]. It was reported that this decline in mobility generally takes place because of structural flaws/defects [21-23].

Several studies have focused on the thermoelectric properties of single layer and few layer $MoS_2$ [12,24-28]. The efficiency of a thermoelectric material could be quantified by figure of merit,

$$ZT = \frac{S^2 \sigma T}{k_{ph} + k_e}$$, where, S is the Seebeck coefficient, σ is electrical conductivity, T is temperature

and $k_e$ and $k_{ph}$ electronic thermal conductivity and lattice thermal conductivity respectively. Since all these parameters are interdependent, enhancing the figure of merit (ZT) is a challenging task. The room temperature figure of merit (ZT) lies between 0.02 to 0.53 [29] for monolayer $MoS_2$. Recent experiments have demonstrated that single layer $MoS_2$ show its potential as a good thermoelectric material with a large value of the Seebeck coefficient [12,28]. According to the formula for figure of merit, ZT, reducing the thermal conductivity is a promising route to improve ZT in low dimensional materials [30,31]. Numerous studies have shown that vacancy, defects, and doping have magnificent effect on thermal conductivity [31-35] in graphene and $MoS_2$; thus influencing the ZT.

Inspired by experimental and theoretical studies we focus in this paper on understanding the role of defects in modulating the electronic and other properties of $MoS_2$ which are of significant impact on their increasing use in device applications, we look at the influence of four different types of atomic defects viz. $V_S$ (one S vacancy), $V_{Mo}$ (one Mo vacancy), $V_{MoS}$ (one Mo and one S vacancy), $V_{MoS_2}$ (one Mo and two S vacancies) on structural, electronic, magnetic and thermoelectric properties of $MoS_2$.

## 2. Computational Details

The pristine $MoS_2$ monolayer has been modeled by 4x4 supercell having 48 atoms. A sufficient vacuum of ~20 Å along the *z*-direction was used to minimize interactions between 2D periodic images. First principle spin polarized calculations, implemented in the Vienna *ab initio* simulation package (VASP) [36,37], have been carried out within the framework of density functional theory (DFT). The projector augmented-wave (PAW) method [38] was employed to represent the electron-ion interactions. The generalized gradient approximation (GGA) using Perdew-Burke-Ernzerhof (PBE) functional has been used to treat the exchange and correlation interactions [39]. The plane wave cutoff energy has been set to 400 eV. The Monkhorst-Pack scheme with a *k* mesh of 10x10x1 has been used to sample the irreducible part of Brillouin zone.

The structural relaxations have been carried out using conjugate gradient technique. All the structures have been fully relaxed until the residual force on each atom is converged to less than 0.02 eV/Å. The electronic density of states (DOS) was calculated with Gaussian broadening parameter of 0.02 eV. The energy convergence criterion has been set to $10^{-6}$ eV self-consistency for electronic steps. The transport coefficients; Seebeck coefficient (S), electrical conductivity (σ) and electronic thermal conductivity ($k_e$) were obtained by semi-classical Boltzmann theory as implemented Boltztrap code [40].

## 3. Results and Discussions

Our investigations begin with the structural optimization of 4x4 supercell of pristine $MoS_2$ with cell dimensions 12.72 Å x 12.72 Å. Figure 1 depicts fully relaxed structure of pristine $MoS_2$ and $MoS_2$ containing atomic defects. The introduction of such atomic defects are very likely in experiments during synthesis and is expected to influence electronic, magnetic and thermoelectric properties of 2D $MoS_2$ as compared to its pristine form.

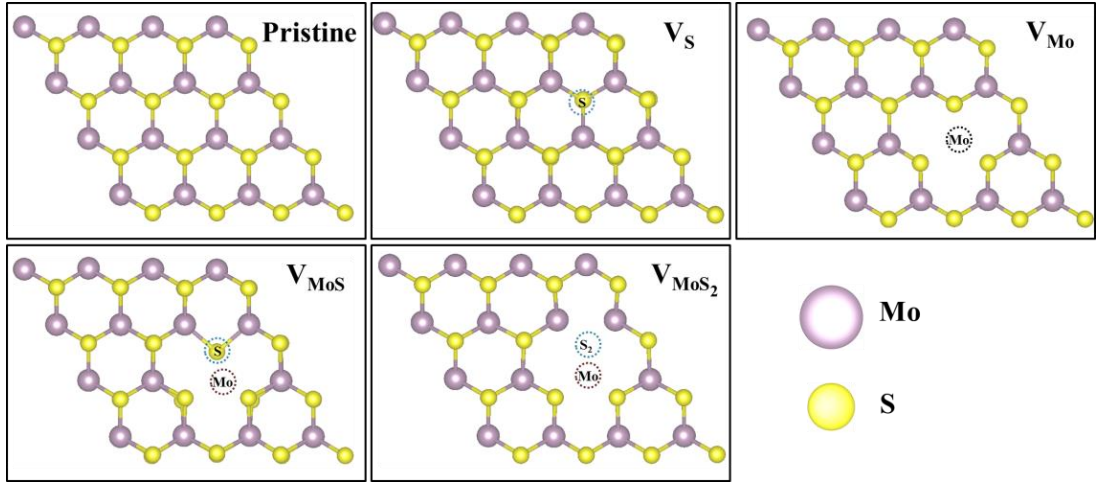

**Figure 1.** Optimized structures of monolayer $MoS_2$ with no vacancy (Pristine), with Mo vacancy ($V_{Mo}$), with MoS vacancy ($V_{MoS}$) and $MoS_2$ vacancy ($V_{MoS2}$).

The formation energy of atomic defects ($E_f$) has been calculated using following equation.

$$E_f = (E_{defect-monolayer} + E_{atomic-defect}) - E_{pristine-MoS_2}$$

Here, $E_{defective-monolayer}$ and $E_{atomic-defect}$ represent the total energy of monolayer $MoS_2$ with atomic defects and total energy of removed atomic defect respectively. The calculated defect formation energy is found to be 5.81 eV for $V_S$ type vacancy which is much lower than the formation energy of all other types of atomic defects under investigations (supplementary table S1). Our calculated formation energy for the case of $V_S$ and $V_{Mo}$ is in close agreement with the earlier reported values of 5.72 eV and 13.5 eV respectively [41,42]. The lower formation energy for the case of $V_S$ type vacancy indicates the ease of formation of such defects among all other types. Note that the positive $E_f$ indicates the endothermic nature of such defects.

In order to find the stability of $MoS_2$ with the presence of defects total energy per atom of pristine and $MoS_2$ with defects is calculated ($E_{total}/n$ ; where $n$ is total number of atoms in supercell). Total energy per atom for pristine and $MoS_2$ with S vacancy is ~0.2 eV lower (more negative) than other types of defects (supplementary Table S1) indicating pristine MoS$_2$ and MoS$_2$ with S defect more stable as compared to other cases.

Next, we pay our attention to the structural distortions taking place due to presence of atomic defects in MoS$_2$. The careful analysis of the relaxed atomic structure we find that structural distortions takes place around the defect only. We find that Mo-Mo distance ($d_{Mo-Mo}$) decreases by 2.66 % and 11.45 % for the V$_{Mo}$ and $V_{MoS_2}$ vacancy, while for V$_S$ and $V_{MoS}$ type of vacancy atomic distortions are of the order of 0.05 Å as compared to pristine MoS$_2$. These atomic distortions are expected to result in different atomic hybridizations which can modulate the electronic structure and magnetic properties of pristine MoS$_2$.

### 3.1 Electronic and Magnetic Properties

Figure 2 show the electronic band structure of pristine and defective MoS$_2$. In the previous investigations electronic band structure of pristine MoS$_2$ shows direct band gap character which is confirmed by PL spectra in experiments [14,43]. Our result is consistent with the earlier reports [14,43,44].

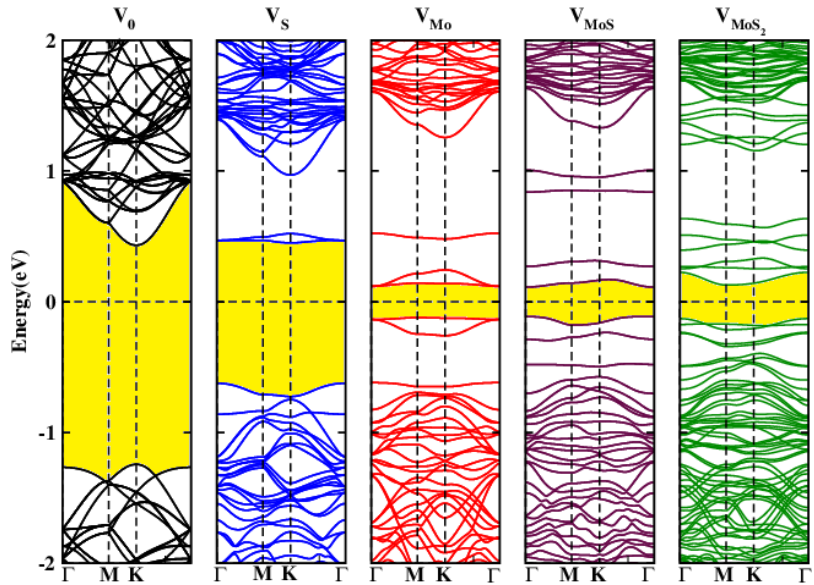

**Figure 2.** Spin Polarized band structures of pristine MoS$_2$ and MoS$_2$ containing different atomic defects. Shaded region indicated the band gap. Fermi level is set at 0 eV.

The electronic band structure shows tuneability with the introduction of defects. The direct band gap of 1.60 eV decreases in defective monolayer. The band gap is found to be 1.07 eV for $V_S$ type defect while band gap reduces to ~ 0.22 eV for all other types of defects. A direct to indirect band gap transition has been observed for defective $MoS_2$ due to appearance of defect energy levels in the vicinity of Fermi level. To gain an insight into localization of Valance Band Maxima (VBM) and Conduction Band Minima (CBM) we have calculated the spin resolved atom projected band structure which is presented in figure 3. It can be seen from figure 3 that the maximum contribution to the VBM and CBM originates from the Mo atom. A similar feature could be seen for the $V_S$ type of defect. For the case of $V_{Mo}$ and $V_{MoS}$ , the defect energy states in the valance band arises from S atom while CBM show contribution from both Mo and S atom. Note that for $V_{MoS_2}$ type defect spin up VBM and CBM are localized on both Mo and S atom while for spin down polarization VBM and CBM are localized on Mo and S atom respectively.

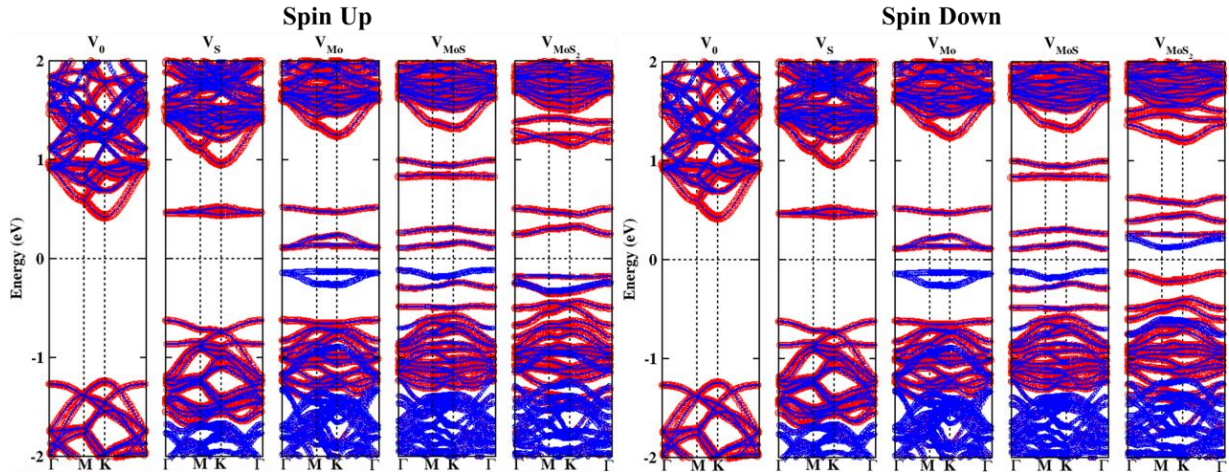

**Figure 3.** Spin resolved atom projected band structure for pristine $MoS_2$ and $MoS_2$ with atomic defects. Red (Blue) color bands are energy bands due to Mo (S) atom respectively. The width of the band represents the spectral weight.

The change in electronic band structures due to atomic defects could be understood by examining the charge density difference profile (defined by, $\Delta\rho = (\rho_{defective-monolayer} + \rho_{atomic-defect}) - \rho_{pristine-MoS_2}$). It can be seen from figure 4 that charge redistribution is more pronounced in the defect region. For the case of $V_{MoS_2}$ the charge accumulates between Mo atoms indicating some sort of strong interaction between Mo atoms. This could be attributed to the structural distortion taking place due to defects.

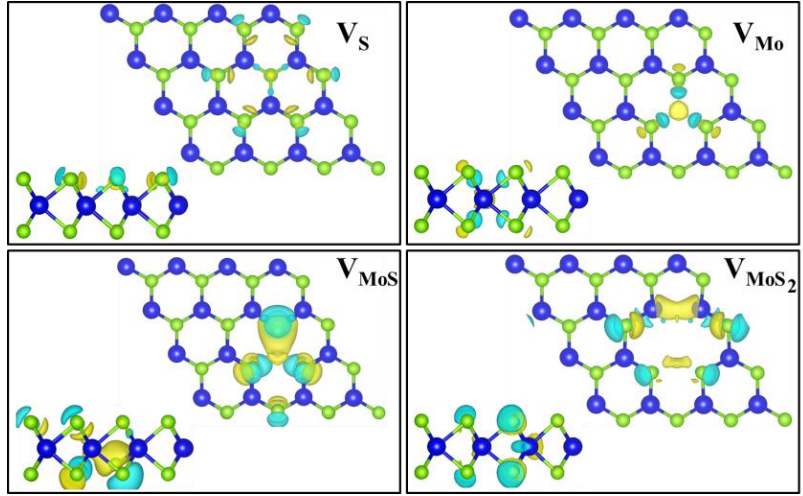

**Figure 4.** Top and side view of charge density difference profiles for S vacancy ($V_S$), Mo vacancy ($V_{Mo}$), MoS Vacancy ($V_{MoS}$) and $MoS_2$ Vacancy ($V_{MoS_2}$). Blue and green balls represent Mo and S atoms respectively. Yellow/Cyan color depicts charge accumulation/depletion. Isosurface value is set to 0.02 e/Å$^3$.

In order to get further insight into the presence of defect states we have analyzed spin polarized density of states (figure 5). The finite density of states in the vicinity of Fermi level for $MoS_2$ with defects clearly depicts the presence of defect states. The presence of defect energy levels are further confirmed by spin resolved atom projected density of states (PDOS) (Supplementary Figure S1). PDOS suggests that the contribution to the valance band in the vicinity of Fermi level is mainly due to Mo-d states in pristine $MoS_2$. In case of $MoS_2$ containing atomic defects, contribution to valance band arises from *S-p* orbitals while conduction band is due to both *Mo-d* and *S-p* orbitals for both the spins. For $V_{MoS_2}$ type of defects Mo-d states contributes to the VBM while S-p (down spin) states contributes to CBM.

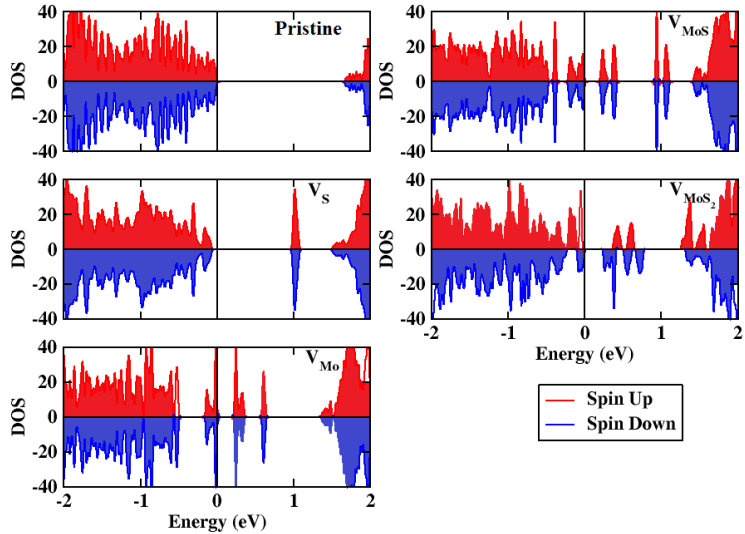

**Figure 5.** Spin polarized total density of states for atomic vacancies in monolayer MoS$_2$.

Next, we turn to determine the magnetic properties of MoS$_2$ with defects. We find that the atomic defects in MoS$_2$ do not lead to the magnetization in the system. Our result for the $V_S$ is in agreement with the conclusions drawn by Jinhua Hong *et. al.* [45]. We find an induced magnetic moment of 1.88 $\mu_B$ for $V_{MoS_2}$. This induced magnetic moment for $V_{MoS_2}$ can be attributed to asymmetry of spin up and spin down density of states (figure 5). Note that for rest of the vacancy types spin up and spin down density of states show symmetry, suggesting that electrons might have occupied same number of spin up and spin down states resulting into no net magnetization. Also, spin density difference plot for the case of $V_{MoS_2}$ (figure S2) suggest that maximum contribution to magnetic moment comes from Mo atoms which is consistent with the PDOS. Also, only nearest neighbor to vacancy site has asymmetric spin densities which confirm the presence of magnetic moment in the system.

## 3.2 Transport Properties

In monolayer with defects, the presence of defect energy levels in the vicinity of Fermi level is expected to modify electron transport properties as compared to the pristine monolayer. The expected modulation in conductance can be quantified by calculating the current-voltage characteristics using a model setup to mimic the scanning tunneling microscope (STM) measurements [46]. In this model set up a ferromagnetic cage like $Fe_{13}$ cluster has been used to simulate the spin polarized probe tip of STM-like setup. The Bardeen, Tersoff and Hamann (BTH) formalism [47,48] has been used to simulate tunneling characteristics for the considered system. The biasing is defined to be forward bias (or positive bias) when the sample is connected to the positive potential with electrons flowing from tip to sample. The tip sample distance is kept at 4Å to ensure the nonbonding configuration between tip and sample. It is worth mentioning here that the magnitude of current exponentially depends upon the tip-sample separation, although the tunneling characteristics remain the same.

The calculated spin resolved tunneling characteristics of the pristine and defective monolayers have been plotted in figure 6 between bias range of -0.5 V to +0.5 V. As the tunneling current is directly proportional to the convolution of DOS between the tip and sample, the finite spin-up and spin-down DOS in the vicinity of the Fermi level of defective monolayer is in principle is the reason of increase in tunneling of both spin carriers (spin-up and spin-down) with the bias voltage. We find that magnitude of current increases at low bias (at ±0.1 V) for both spin carriers which can be attributed to the presence of defect energy levels in defective monolayers. Interestingly, for $V_{MoS_2}$ case magnitude of current due to spin up carriers modulates at higher bias as compared to spin down case. This could be attributed to difference in available spin up and spin down channels in the vicinity of Fermi level (figure 3). This interesting feature underlines the potential of $MoS_2$ with $V_{MoS_2}$ type defects in spin based electronic devices.

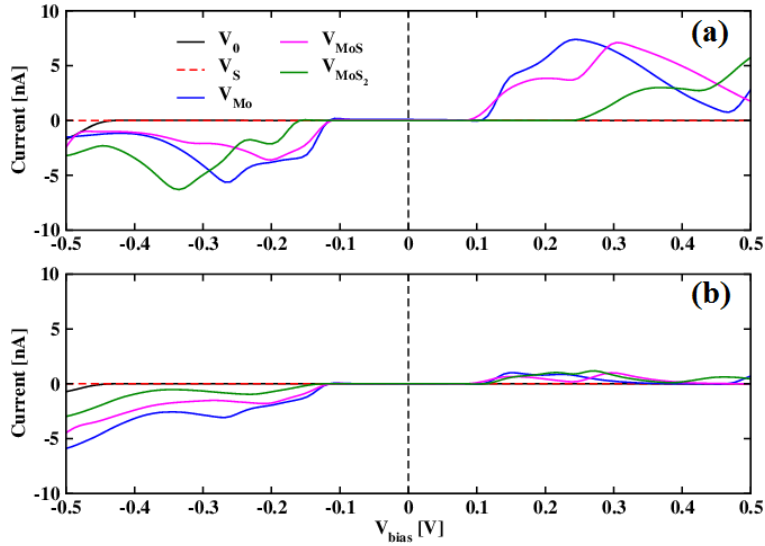

**Figure 6:** Spin polarized I-V Characteristics of pristine and defective MoS₂ monolayer; a) spin up, b) spin down.

### 3.3 Thermoelectric Properties

Our analysis suggests that MoS$_2$ monolayer with defects exhibit drastic change in electronic structures around Fermi level from that of pristine form. MoS$_2$ monolayer with atomic defects has strongly reduced band gap and an enhanced density of states around the Fermi level. Let us now examine how the thermoelectric properties of defective MoS$_2$ monolayer differ from that of pristine counterpart which has not been reported yet. The new energy levels in defective MoS$_2$ play a crucial role in determining the thermoelectric performance of material. A good thermoelectric material must be able to convert dissipating heat into electrical energy efficiently. The modulation in electronic properties of pristine MoS$_2$ with atomic defects is pointing to an expected influence on thermoelectric performance of MoS$_2$. Figure S3 − S5 (supplementary information) summarizes the calculated S, σ, σ S$^2$ as a function of chemical potential (μ) at different temperatures. In these calculations chemical potential is positive when Fermi level is raised (lowered), which corresponds to n-type (p-type) doping. The Seebeck coefficients are lower for V$_{Mo}$ and V$_{MoS}$ relative to pristine MoS$_2$. The maximum room temperature Seebeck coefficient is ~1000 and (~500) μV/K for V$_S$ and $V_{MoS_2}$ respectively (Figure S3). The large Seebeck coefficients $V_S$ and $V_{MoS_2}$ is related to the fact that MoS$_2$ with such defects introduces new energy levels around the Fermi level leading to asymmetric density of states between valance band and conduction band resulting in high Seebeck coefficient. The Seebeck coefficient decreases with the increase in temperature, however, the values of S are comparable to earlier reported values at 300K [29].

Since the efficiency of thermoelectric materials could be enhanced by increasing Power Factor ($\sigma S^2$). Despite the fact that the Seebeck coefficients for $V_S$ and $V_{MoS_2}$ are higher than the pristine $MoS_2$ and other cases, the conductivity ($\sigma$) has also a strong effect in controlling the Power Factor than the Seebeck coefficient. An enhancement in $\sigma$ is observed for p-type doping for pristine and $V_S$ case while for all other cases '$\sigma$' continuously modulates with chemical potential (Figure S3 and S4). Note that relative to pristine $MoS_2$, $MoS_2$ with defects show decreased conductivity ($\sigma$).

The figure of merit (ZT) is a key parameter to quantify the efficiency of thermoelectric materials and devices. The optimization of ZT value also depends on the electronic thermal conductivity ($k_e$) and phonon thermal conductivity ($k_{ph}$). The theoretical [26,49] and experimental [27,50] literature reports phonon thermal conductivity of monolayer and few-layer $MoS_2$ with values ranging from 0.24 $Wm^{-1}K^{-1}$ to 116.8 $Wm^{-1}K^{-1}$. Here we employed the experimental value of $k_{ph}$= 35.4 $Wm^{-1}K^{-1}$ by R. Yan *et. al.* [27] along with our calculated electronic thermal conductivity to calculate total thermal conductivity ($k = k_{ph} + k_e$). Figure S6 depicts calculated total thermal conductivity as a function of chemical potential at different temperatures for pristine and defective monolayer $MoS_2$. A significant reduction (10 times) in thermal conductivity has been found for the $MoS_2$ with defects. Relative to different types of considered atomic defects, $V_S$ and $V_{MoS_2}$ show highly reduced total thermal conductivity pointing towards an enhanced ZT.

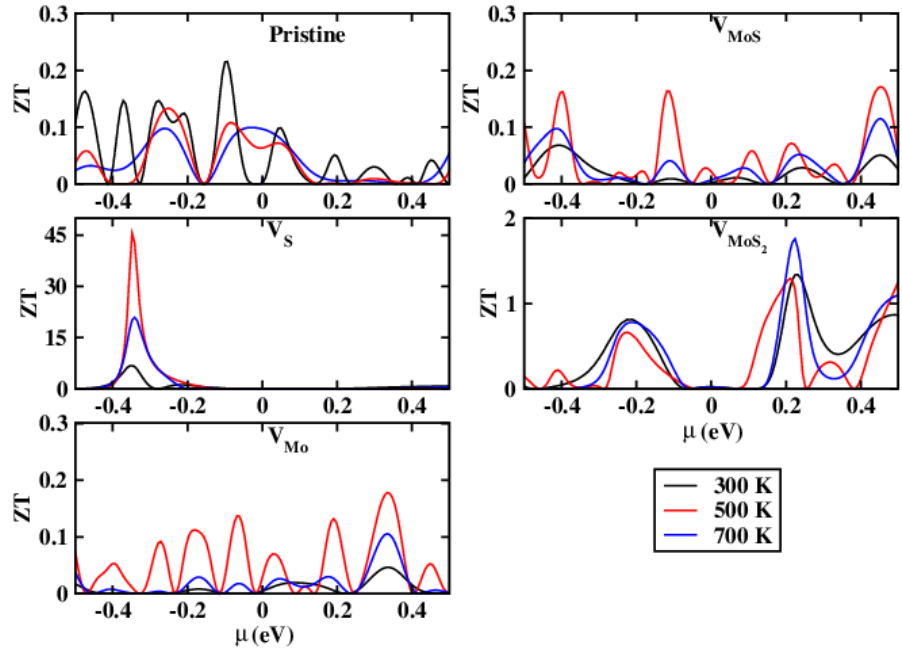

**Figure 7:** Calculated Figure of merit (ZT) for different atomic defects as a function of chemical potential at 300 K, 500 K and 700 K.

Figure 7 summarizes the figure of merit dependence on the chemical potential (μ) at 300, 500 and 700 K. The highest value of ZT is found to be 0.21 at 300 K which in agreement with the earlier reported values ranging between 0.02 to 0.53 [29] for pristine MoS$_2$. The value of ZT modulates around 0.2 for the MoS$_2$ with defects. A room temperature (at 300 K) ZT of 6.24 and 1.30 has been calculated for $V_S$ and $V_{MoS_2}$. An enhanced ZT for V$_S$ and $V_{MoS_2}$ is attributed to higher (lower) Seebeck (electronic thermal conductivity). The thermoelectric properties also show tuneability with the temperature. Therefore, in nutshell, our results suggest that the atomic defects and temperature show enhancing influence on the thermoelectric properties of MoS$_2$ monolayer and possibly other layered TMDs (MX$_2$; M=Mo, W; X=S, Se).

**Conclusions**

A first principle study has been carried out to study the electronic and thermoelectric properties of MoS$_2$ monolayer with atomic defects. Following can be concluded from this study:

- The defect formation energy for Sulphur defect is found to be +5.81 eV which is much lower than that found for other atomic defect cases.

- The reduction of interatomic distance by 2.66 % and 11.45 % around the defect sites is significant for $V_{Mo}$ and $V_{MoS_2}$ type defects, while for $V_S$ and $V_{MoS}$ type of vacancy atomic distortions are of the order of 0.05 Å.

- The electronic band gap shows tuneability with the atomic defects. Band gap gets reduced to 0.22 eV with a direct to indirect band gap transition due to appearance of defect energy levels.

- An induced magnetic moment of 1.88 $\mu_B$ has been found for $V_{MoS_2}$ only.

- The magnitude of current increases at low bias (at ±0.1 V) for both spin carriers suggesting presence of defect energy levels in monolayers with defects.

- The thermoelectric properties including Seebeck coefficient, electrical and thermal conductivity, power factor and figure of merit show tuneability with temperature. The room temperature ZT of $MoS_2$ with S and $MoS_2$-type defects can reach as high as 6.24 and 1.30 respectively.

Thus, our investigations have systematically provided a significant understanding of the variation in electronic, magnetic, transport and thermoelectric properties induced by atomic defects which can be used as a reference point for exploring potential applications at nanoscale.

**Acknowledgements**

Munish Sharma wishes to acknowledge the DST, Govt. of India, New Delhi for providing the financial support in the form of INSPIRE Fellowship. CVRAMAN, high performance computing cluster (provided by FIST, DST, Govt. of India, New Delhi) at Physics Department, Himachal Pradesh University and K2 high performance computing cluster at IUAC have been used to obtain results presented in this paper.

**Supplementary Information**

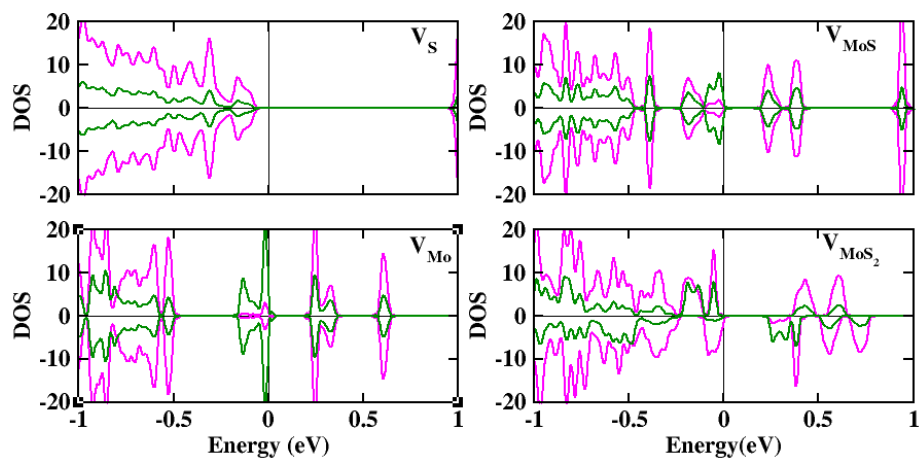

**Figure S1.** Spin polarized atom projected density of states for atomic vacancies in monolayer MoS$_2$. Magenta and Green color represents Mo-4d and S-3p orbital DOS respectively.

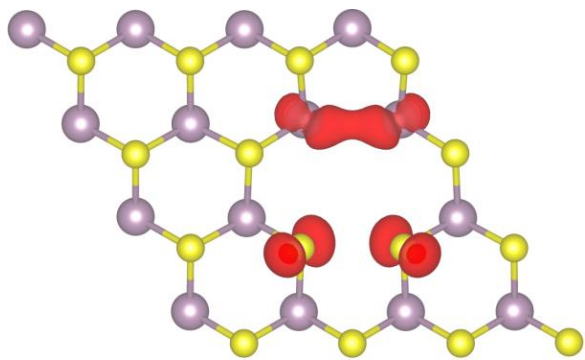

**Figure S2.** The spin density difference plot (spin up - spin down) for $V_{MoS_2}$. Red color represents spin up density. Isosurface values are set at 0.005 e/(Å)$^3$

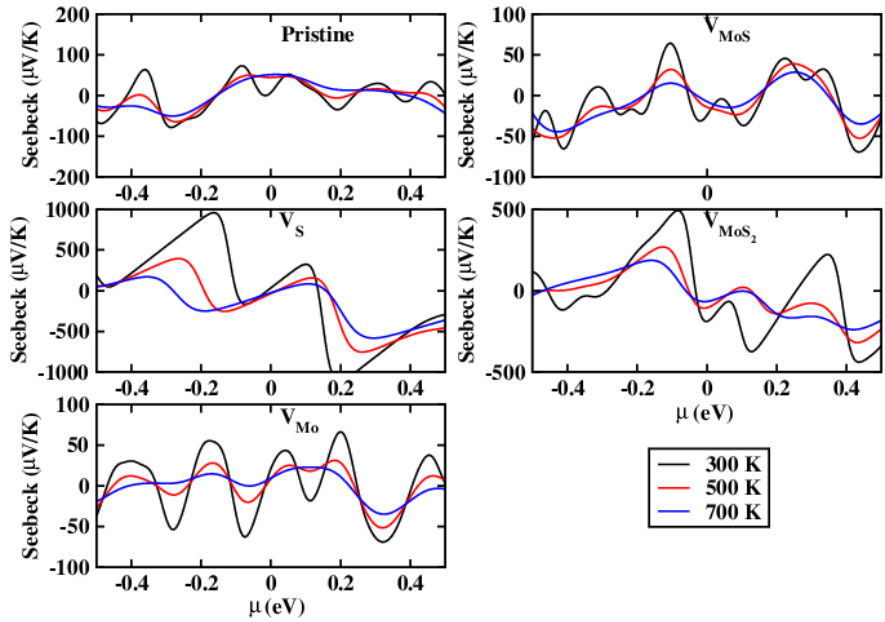

**Figure S3:** Calculated Seebeck coefficient for different atomic vacancies as a function of chemical potential.

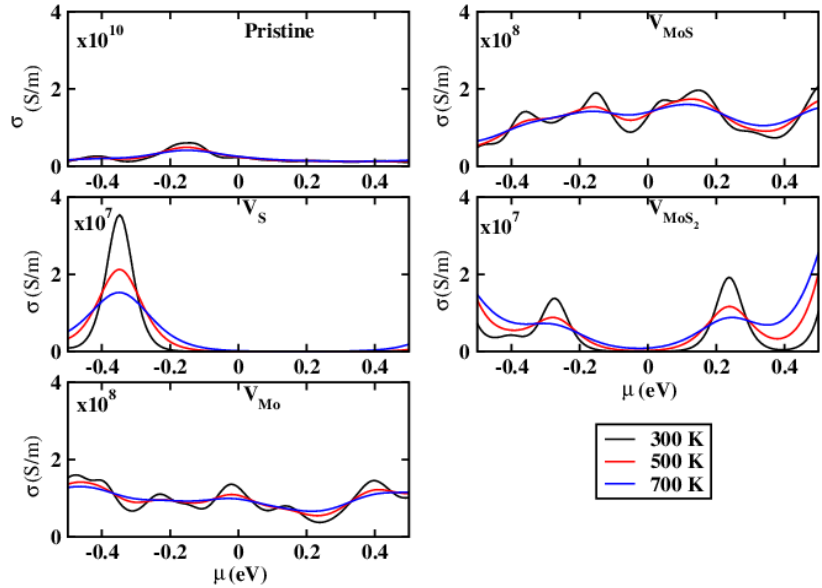

**Figure S4:** Calculated electrical conductivity (σ) for different atomic vacancies as a function of chemical potential.

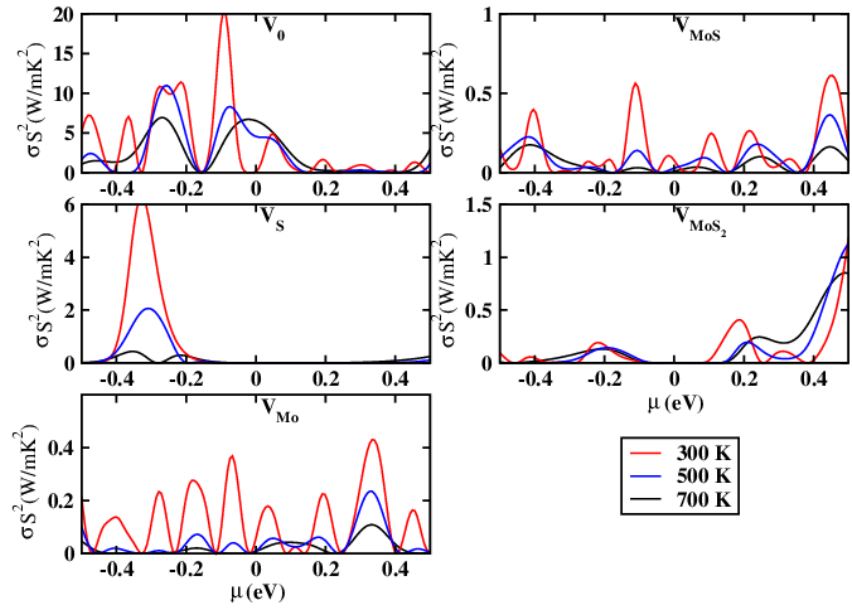

**Figure S5:** Calculated Power Factor ($\sigma S^2$) for different atomic vacancies as a function of chemical potential.

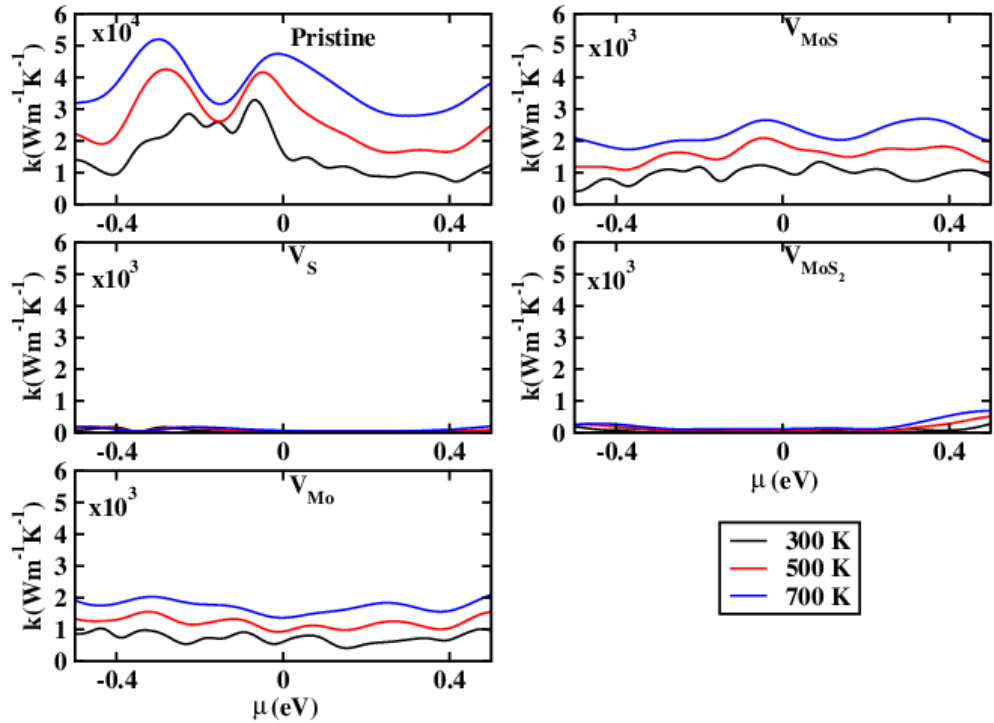

**Figure S6:** Calculated thermal conductivity ($k = k_{ph} + k_e$) as a function of chemical potential at different temperatures for pristine and defective monolayer MoS$_2$. Here k$_{ph}$ = 35.4 Wm$^{-1}$K$^{-1}$ [1]

**Table S1:** Total energy per atom and atomic defect formation energy for pristine and defective $MoS_2$.

| System | Total Energy Per atom (eV) | Formation energy (eV) |
|---|---|---|
| **Pristine MoS$_2$** | -7.26 | - |
| **V$_S$** | -7.27 | +5.81, 5.72 [2] |
| **V$_{Mo}$** | -7.03 | +13.56, 13.5 [3] |
| **V$_{MoS}$** | -7.07 | +13.57 |
| $V_{MoS_2}$ | -7.09 | +15.57 |